\documentclass[usenatbib]{mn2e}

\usepackage{epsfig}
\usepackage{mathrsfs,aas_macros}

\def\Lya{Ly$\alpha$~} 
\def\Lyb{Ly$\beta$~}

\def\HI{\hbox{H$\,\rm \scriptstyle I\ $}}
\def\HII{\hbox{H$\,\rm \scriptstyle II\ $}}

\def\NV{\hbox{N$\,\rm \scriptstyle V\ $}} 

\title[How neutral is the IGM around ULAS J1120$+$0641?]{How neutral is the intergalactic medium surrounding the
  redshift \boldmath{$z=7.085$} quasar ULAS J1120\boldmath{$+$}0641?}

\author[J.S. Bolton et al.] {J.S. Bolton$^{1}$, M.G. Haehnelt$^{2}$,
    S.J. Warren$^{3}$, P.C. Hewett$^{2}$, D.J. Mortlock$^{3}$,
  \newauthor B.P. Venemans$^{4}$, R.G. McMahon$^{2}$ \& C. Simpson$^{5}$
  \\ $^1$ School of Physics, University of Melbourne, Parkville, VIC
  3010, Australia \\ $^2$ Kavli Institute for Cosmology and Institute
  of Astronomy, Madingley Road, Cambridge, CB3 0HA \\ $^3$
  Astrophysics Group, Imperial College London, Blackett Laboratory,
  Prince Consort Road, London, SW7 2AZ \\ $^4$ European Southern
  Observatory, Karl-Schwarzschild Strasse 2, 85748 Garching bei
  M\"unchen, Germany \\ $^5$ Astrophysics Research Institute,
  Liverpool John Moores University, Twelve Quays House, Egerton Wharf,
  Birkenhead, CH41 1LD}

\begin{document}

\date{}

\maketitle

\label{firstpage}

\begin{abstract}
The quasar ULAS J1120$+$0641 at redshift $z=7.085$ has a highly
ionised near zone which is smaller than those around quasars of
similar luminosity at $z\simeq 6$. The spectrum also exhibits evidence
for a damping wing extending redward of the systemic \Lya redshift.
We use radiative transfer simulations in a cosmological context to
investigate the implications for the ionisation state of the
inhomogeneous IGM surrounding this quasar.  Our simulations show that
the transmission profile is consistent with an IGM in the vicinity of
the quasar with a volume averaged \HI fraction of $\langle f_{\rm
  HI}\rangle_{\rm V}\ga 0.1$ and that ULAS J1120$+$0641 has been
bright for $10^{6}$--$10^{7}\rm\,yr$. The observed spectrum is also
consistent with smaller IGM neutral fractions, $\langle f_{\rm
  HI}\rangle_{\rm V}\sim 10^{-3}$--$10^{-4}$, if a damped \Lya system
in an otherwise highly ionised IGM lies within $5$ proper Mpc of the
quasar.  This is, however, predicted to occur in only $\sim 5$ per
cent of our simulated sight-lines for a bright phase of
$10^{6}$--$10^{7}$ yr.  Unless ULAS J1120$+$0641 grows during a
previous optically obscured phase, the low age inferred for the quasar
adds to the theoretical challenge of forming a $2\times
10^{9}\rm\,M_{\odot}$ black hole at this high redshift.

\end{abstract}
 


\section{Introduction}
 
The spectra of quasars with redshifts $z\ga 6$ are an important probe
of the ionisation state of the intergalactic medium (IGM) in the early
Universe.  The intergalactic \Lya opacity rapidly rises with
increasing redshift, and consequently there has been much debate
whether these quasars have been observed before the Universe was fully
reionised ({\it e.g.} \citealt{Becker07,Lidz07,Mesinger10}).  Studies
of the small windows of \Lya transmission observed in close proximity
to these objects provide important clues in this respect
(\citealt{WyitheLoeb04,MesingerHaiman04,Fan06,BoltonHaehnelt07,AlvarezAbel07,Maselli09}).
These highly ionised ``near zones'' lie between the red edge of the
\cite{GunnPeterson65} trough and the quasar \Lya emission line, and
arise because the hydrogen close to these sources is highly ionised.

\cite{Mortlock11} (hereafter M11) have recently reported the highest
redshift quasar found to date, ULAS J1120$+$0641 at $z=7.085$,
discovered using data from the UKIRT Infrared Deep Sky Survey (UKIDSS,
\citealt{Lawrence07}).  In their analysis of the \Lya near zone
observed in a low resolution ($R\simeq 1400$) VLT/FORS2 spectrum of
moderate signal-to-noise, M11 find the extent and shape of the near
zone is distinct from those around the much debated $z\sim 6$ quasars.
Assuming the quasar is embedded in a homogeneous IGM, M11 concluded
the observed transmission profile of ULAS J1120$+$0641 is consistent
with a surrounding IGM for which the volume averaged neutral fraction
is $\langle f_{\rm HI}\rangle_{\rm V}>0.1$ at $z\simeq 7.1$.

In this Letter we examine the implications of the M11 observation in
more detail using simulations of radiative transfer through an
inhomogeneous IGM.  These simulations have successfully explained the
properties of \Lya near zones observed at $z\simeq 6$ in the context
of a highly ionised IGM (\citealt{Bolton10}).  We begin in Section 2
by giving a brief overview of \Lya near zone sizes.  In Section 3 we
describe our numerical simulations before comparing them directly to
the M11 measurement in Section 4.  Finally in Section 5 we discuss the
implications of our results and conclude.  We assume the cosmological
parameters $\Omega_{\rm m}=0.26$, $\Omega_{\Lambda}=0.74$,
$\Omega_{\rm b}h^{2}=0.023$ and $h=0.72$ throughout, and unless
otherwise stated all distances are given in proper units.


\vspace{-0.4cm}
\section{Quasar near zone sizes}

\begin{figure}
\begin{center}
  \includegraphics[width=0.45\textwidth]{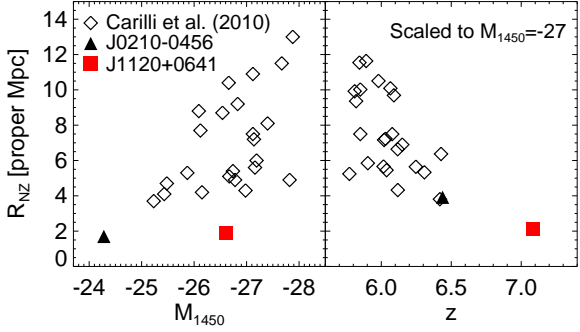}
\vspace{-0.3cm}
\caption{Summary of near zone size measurements in the current
  literature.  {\it Left:} Near zone sizes against the quasar absolute
  magnitude. {\it Right:} Near zone sizes against redshift. The sizes
  have been rescaled by a factor $10^{0.4(27 + M_{1450})/3}$
  \citep{Fan06} to correspond to a common absolute magnitude of
  $M_{1450}=-27$.  Note this scaling assumes $R_{\rm NZ} \propto {\dot
    N}^{1/3}$, where ${\dot N}$ is the emission rate of ionising
  photons from the quasar, appropriate if the near-zone traces the
  boundary of the quasar \HII region. If the near-zone extent is
  instead set by resonant absorption, $R_{\rm NZ}\propto {\dot
    N}^{1/2}$ (\citealt{BoltonHaehnelt07}).}
  \label{fig:obs_compiled}
\end{center}
\end{figure}

At the luminosities typical of observed $z\simeq 6$ quasars, and
assuming an optically bright phase of $t_{\rm q}\sim 10^{7}\rm\,yr$,
the size of a quasar \HII region expanding into a partially neutral
IGM and that of a proximity zone embedded in an already highly ionised
IGM are comparable (\citealt{BoltonHaehnelt07}).  Further, if the
surrounding IGM is significantly neutral, the red Gunn--Peterson (GP)
damping wing will reduce the observed size of the near zone and \Lya
absorption may extend redward of the quasar systemic redshift
(\citealt{MiraldaRees98,MesingerHaiman04}).  Unfortunately, the
possibility of confusing this damping wing with a collapsed region
with a high \HI column density in an otherwise highly ionised IGM
complicates matters further.  These issues make the interpretation of
the \Lya near zones observed around these quasars with respect to the
IGM ionisation state ambiguous
(\citealt{BoltonHaehnelt07,Maselli07,Lidz07}).

A summary of the existing measurements of \Lya near zone sizes around
high redshift quasars, compiled recently by \cite{Carilli10}, is
displayed in Fig.~\ref{fig:obs_compiled}.  Interestingly, ULAS
J1120$+$0641 has a small\footnote{The only quasar with a similarly
  small near zone is CFHQS J0210$-$0456 at $z=6.44$
  (\citealt{Willott10}), but this object is considerably fainter and
  is consistent with the other $z\simeq 6$ quasars on accounting for
  its fainter luminosity.} near zone size given its bright absolute AB
magnitude, $M_{1450}=-26.6$.  This is demonstrated by the evolution of
observed near zone sizes with redshift, rescaled to correspond to a
common magnitude of $M_{1450}=-27$, in the right-hand panel of
Fig.~\ref{fig:obs_compiled}.  M11 have noted that the small near zone
size around ULAS J1120$+$0641, coupled with \Lya absorption redward of
the systemic redshift from a putative red damping wing, may indicate
the IGM surrounding ULAS J1120$+$0641 is significantly more neutral
than around quasars at $z\sim 6$. We now investigate this in more
detail using realistic synthetic absorption spectra which employ a
variety of evolutionary histories for the IGM ionisation state.

\begin{figure}
\begin{center}
  \includegraphics[width=0.45\textwidth]{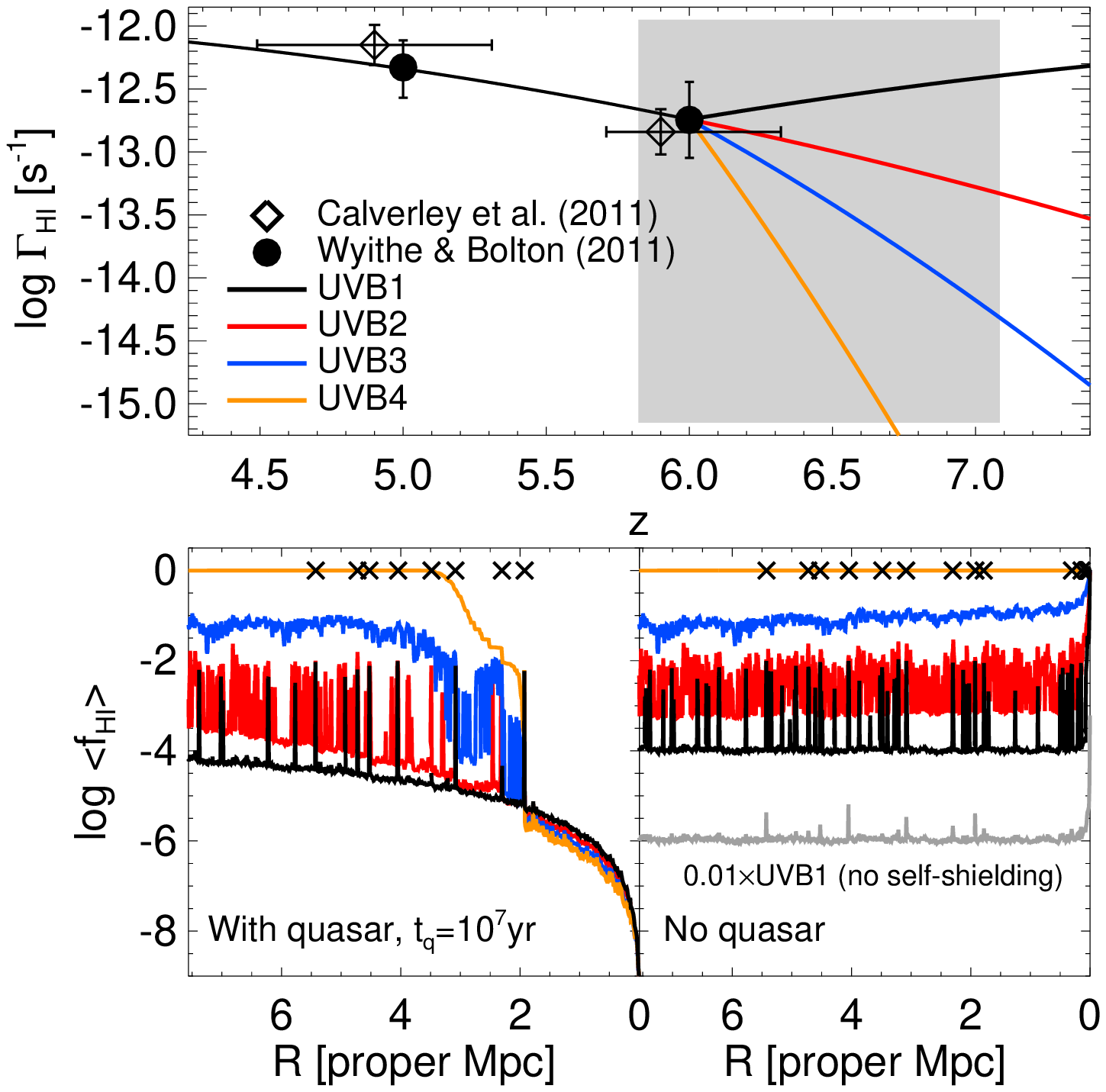}
  \vspace{-0.2cm}
  \caption{{\it Top:} Four simple models for the redshift evolution of
    the background \HI photo-ionisation rate along a quasar
    sight-line.  The grey shading represents the range covered by the
    \Lya forest to rest frame \Lyb for a quasar at $z=7.085$.  The
    measurements were obtained by \citet{WyitheBolton11} from the \Lya
    forest opacity and by \citet{Calverley11} using the proximity
    effect. Note that for UVB4 we set $\Gamma_{\rm HI}=0$ at
    $z>6.9$. {\it Bottom:} The neutral hydrogen fraction averaged over
    100 simulated sight-lines produced by the UVB models including
    (left) and excluding the quasar (right).  The averaging is
    performed for display purposes only, and results in neutral
    regions being displayed as spikes with $\langle f_{\rm HI}\rangle
    \sim10^{-2}$. The crosses mark the position of all DLAs in the
    sight-lines constructed using UVB4, and the grey curve in the
    right panel shows the logarithm of $\langle f_{\rm HI}\rangle$ for
    UVB1 {\it excluding} the self-shielding approximation, multiplied
    by a factor of $0.01$ for clarity.}
  \label{fig:UVB_models}
\end{center}
\end{figure}


\vspace{-0.4cm}
\section{Numerical simulations} \label{sec:sims}

We construct \Lya absorption spectra using a high resolution
hydrodynamical simulation combined with a line-of-sight radiative
transfer code (\citealt{Bolton10}).  The hydrodynamical simulation was
performed using the {\sc GADGET-3} code (\citealt{Springel05}) and has
a box size of $10h^{-1}$ comoving Mpc and a gas particle mass of
$9.2\times 10^{4}h^{-1}M_{\odot}$.  A total of 100 sight-lines of
length $96.6h^{-1}$ comoving Mpc were extracted around haloes
identified in the simulation to construct skewers through the IGM
density, temperature and peculiar velocity field.  The largest halo
mass in the simulation volume at $z=7.1$ is $4.7\times 10^{10}h^{-1}
M_{\odot}$, which likely underestimates the ULAS J1120$+$0641 host
halo mass.  However, resolving the low density IGM is more important
for modelling the \Lya absorption, and a small, high resolution
simulation provides the best compromise (\citealt{Bolton10}).
Following M11 and \cite{Bolton10}, we assume the spectrum of the
unobscured, radio quiet quasar ULAS J1120$+$0641 is a broken power law
where $f_{\nu}\propto \nu^{-0.5}$ for $1050{\rm\,\AA} <\lambda <
1450\,\rm \AA$ and $f_{\nu}\propto \nu^{-1.5}$ for $\lambda<1050\,\rm
\AA$, yielding an ionising photon emission rate of ${\dot N}=1.3\times
10^{57}\rm\,s^{-1}$.

\begin{table*}
  \centering
  \begin{minipage}{180mm}
    \begin{center}
      
      \caption{Average \HI fractions and the incidence of optically
        thick absorbers within $0.05{\rm\,Mpc}<R<5{\rm\,Mpc}$ of the
        quasar host halo in all 100 simulated sight-lines including
        the self-shielding prescription.  The absorber columns are
        estimated by integrating the \HI densities over a scale of
        $20$ kpc.  From left to right, the columns list the volume and
        mass averaged \HI fraction, the number of LLSs
        ($10^{17.2}\rm\,cm^{-2}<N_{\rm HI}<10^{20.3}\rm\,cm^{-2}$) and
        DLAs ($N_{\rm HI}>10^{20.3}\rm\,cm^{-2}$) per unit redshift
        and the fraction of sight-lines where there is at least one
        LLS or DLA, {\it all prior to the quasar turning on}.  In the
        final four columns the latter are also given after a quasar
        with age $10^{6}\rm\,yr$ and $10^{7}\rm\,yr$ has ionised the
        surrounding IGM. Note that for models UVB3 and UVB4, where the
        IGM has a large neutral fraction, identifying LLSs becomes
        ambiguous.  At $z=7.085$, $ \Delta z =1$ corresponds to $\sim
        43.8$ Mpc.}
      
      \vspace{-2mm}
      \begin{tabular}{c|c|c|c|c|c|c|c|c|c|c}
        \hline
        Model  & 
        $\langle f_{\rm HI} \rangle_{\rm V}$ & 
        $\langle f_{\rm HI} \rangle_{\rm M}$ & 
        
        $\frac{dn_{\rm LLS}}{dz}$ & 
        $\frac{dn_{\rm DLA}}{dz}$ &
        
        $f_{\rm los}^{\rm LLS}$  & 
        $f_{\rm los}^{\rm DLA}$ & 
        
        $f_{\rm los}^{\rm LLS}$($10^{6}$yr)& 
        $f_{\rm los}^{\rm DLA}$($10^{6}$yr)& 
        
        $f_{\rm los}^{\rm LLS}$($10^{7}$yr)& 
        $f_{\rm los}^{\rm DLA}$($10^{7}$yr)\\

        \hline
        
        UVB1 & $5.0\times 10^{-4}$ & $2.7\times 10^{-2}$ & $3.7$ & $0.3$ & $0.33$ & $0.03$ 
        & $0.16$ & $0.03$ & $0.05$ & $0.03$  \\
        
        UVB2 & $6.2\times 10^{-3}$ & $8.3\times 10^{-2}$ & $55.4$ & $0.7$ & $0.99$ & $0.08$
        & $0.75$ & $0.07$ & $0.10$ & $0.04$ \\
        
        UVB3 & $1.1\times 10^{-1}$ & $3.4\times 10^{-1}$ & -- & $0.8$ & -- & $0.09$
        & -- & $0.07$ & -- & $0.06$ \\
        
        UVB4 & $1.0$ & $1.0$ & -- & $1.0$ & -- & $0.11$
        & -- & $0.08$ & -- & $0.07$ \\

        \hline
 
      \end{tabular}
    \end{center}
  \end{minipage}
  \label{tab:sims}
\end{table*}

The ionisation state of the IGM along the sight-lines prior to the
quasar turning on was initialised using four different models for the
background \HI photo-ionisation rate, displayed in
Fig.~\ref{fig:UVB_models}.  However, it is important to note that
reionisation is an inhomogeneous process, and as a result each model
may be realised at the same time in different regions of the IGM
(\citealt{MesingerFurlanetto08}).  Furthermore, the ionisation state
of the IGM in the vicinity of a massive host halo will be biased due
the clustering of lower luminosity sources
(\citealt{Lidz07,Wyithe08}).  These models are thus designed to
explore a range of different initial \HI fractions in the IGM close to
the quasar, but {\it are not representative of global reionisation
  histories}.

\vspace{-0.0275cm}

As we only compute the radiative transfer for the quasar ionising
radiation, we also add a simple prescription for regions which are
self-shielded from the ionising background. Assuming that the typical
size of an \HI absorber is the Jeans scale (\citealt{Schaye01}), the
hydrogen number density where an \HI absorber begins to self-shield
can be approximated as $n_{\rm H} \simeq 3.6 \times
10^{-3}{\rm\,cm^{-3}}\,(\Gamma_{\rm
  HI}/10^{-12}\rm\,s^{-1})^{2/3}(T/10^{4}\rm\,K)^{2/15}$.  We use this
expression to compute the density threshold at which hydrogen is
self-shielded {\it prior} to being ionised by the quasar, and set the
background photo-ionisation rate to zero in these regions.  The weak
temperature dependence is computed self-consistently using the gas
temperature in our simulations.  The lower panel of
Fig.~\ref{fig:UVB_models} displays the resulting neutral hydrogen
fraction averaged over all 100 sight-lines assuming $t_{\rm
  q}=10^{7}\rm\,yr$ and including (left panel) and excluding (right
panel) ionising radiation from the quasar.  The grey curve in the
right panel also shows the \HI fraction for UVB1 excluding the
self-shielding model. Self-shielded regions become rapidly more common
for ultra-violet background (UVB) models with decreasing \HI
photo-ionisation rates, increasing the volume averaged neutral
fraction and producing a patchy ionisation structure along each
sight-line.  A summary of the average \HI fractions and incidence of
optically thick absorbers in all models which include the
self-shielding prescription is given in Table 1.  Note that the quasar
photo-ionises optically thick absorbers in its vicinity, significantly
reducing the incidence of these systems.  However, the impact of this
process is reduced for shorter quasar ages or the smaller ionising
photon mean free path associated with a larger IGM neutral fraction.


\vspace{-0.4cm}
\section{Results}

\begin{figure*}
\centering
\begin{minipage}{180mm}
\begin{center}
\psfig{figure=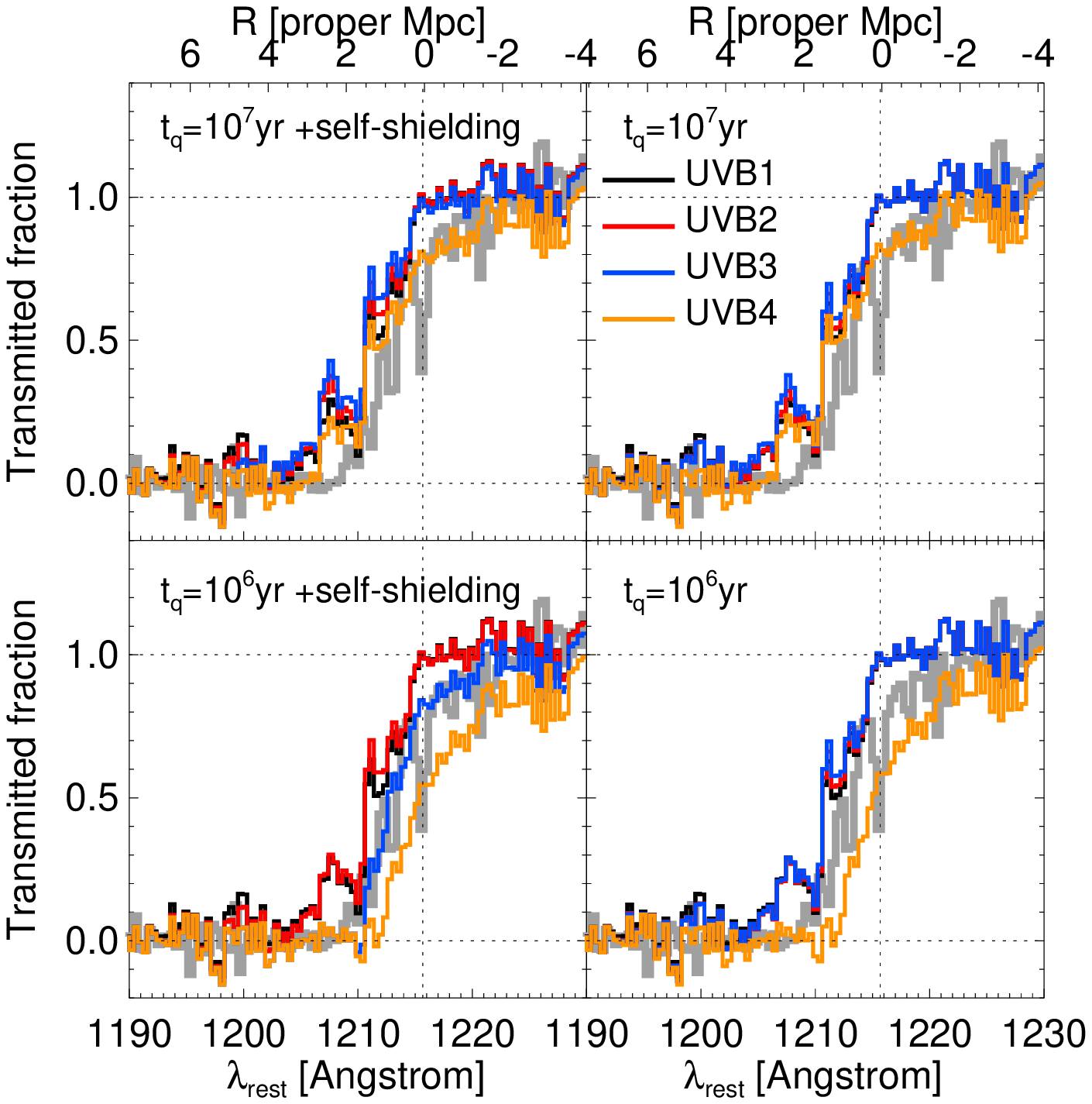,width=0.435\textwidth}
\psfig{figure=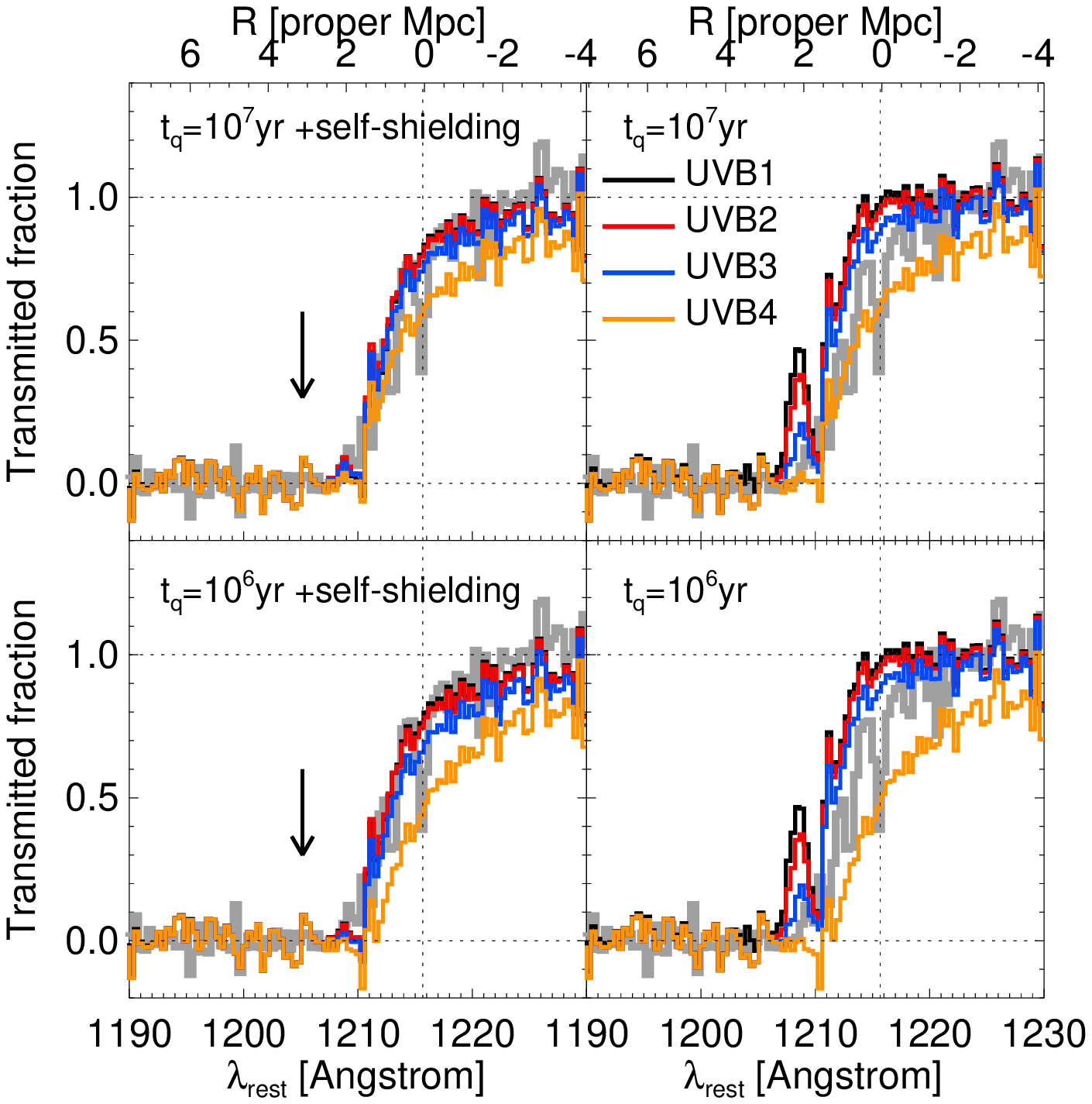,width=0.435\textwidth}

\vspace{-0.3cm}

\caption{Simulated quasar spectra constructed using the four different
  UVB models displayed in Fig.~\ref{fig:UVB_models}. The spectra are
  processed to mimic the M11 spectrum by convolving with a Gaussian
  instrument profile with FWHM$=240\rm\,km\,s^{-1}$, rebinning onto
  pixels of width $0.4\rm\,\AA$ (rest frame) and adding Gaussian
  distributed noise using the observed noise array.  The observed
  transmission of ULAS J1120$+$0641 is shown by the light grey curve in
  each panel.  {\it Left:} Sight-line 1.  Note the impact of the red
  GP damping wing on the spectrum for models with a fully neutral IGM
  (orange curves). {\it Right:} Sight-line 2.  In this sight-line a
  prominent red damping wing is seen for all UVB models when the
  self-shielding prescription is included.  A DLA with $N_{\rm
    HI}\simeq 10^{20.5}\rm\,cm^{-2}$ is present at the position marked
  by the arrow.}
\label{fig:spectra}
\end{center}
\end{minipage}
\end{figure*}

We compare the measured transmission profile around ULAS J1120$+$0641,
described in detail by M11, to two of our simulated sight-lines in
Fig.~\ref{fig:spectra}.  For each sight-line we consider quasar ages
of $10^{6}\rm\,yr$ and $10^{7}\rm\,yr$, and for each quasar age the
effect of optically thick absorbers is illustrated by models which
include and exclude the self-shielding approximation. This qualitative
comparison highlights an important point: for an initially neutral IGM
(UVB4, orange curves) or a partially neutral IGM (UVB3, blue curves)
and a short quasar bright phase ($10^{6}\rm\,yr$) a prominent damping
wing with absorption extending redward of rest frame \Lya is present.
However, optically thick absorbers will also have an impact on the
near zone transmission profile redward of \Lya if a collapsed neutral
region is close to the quasar; the presence of a damped \Lya absorber
(DLA) with $N_{\rm HI}\simeq 10^{20.5}\rm\,cm^{-2}$ at $3.1\rm\,Mpc$,
marked by the arrow in the right main panel, demonstrates this.

We may consider the relative likelihood of these possibilities by
analysing the transmission profiles for all 100 of our simulated
spectra.  Considering the moderate signal-to-noise and low resolution
of the observed spectrum of ULAS J1120$+$0641 and the difficulty of
continuum fitting due to the complexity of the intrinsic \Lya and \NV
emission, we use two simple properties to characterise the
transmission profile for a quantitative comparison. These are: the
distance from the quasar where the transmitted fraction first falls
below 10 per cent, $R_{\rm NZ}$ (\citealt{Fan06}), and the transmitted
fraction at the rest frame $\rm Ly\alpha$ wavelength,
$T_{1216}$. These are measured after smoothing the spectra with a box
car window of width $2.4\rm\,\AA$ (rest frame) to remove fluctuations
on smaller scales due to noise and absorption lines.  The
corresponding values\footnote{The measurements include a systematic
  error due to the uncertain continuum placement (see Figure 4 in
  M11).} for ULAS J1120$+$0641 are $R_{\rm NZ}=1.9\pm 0.1\rm\,Mpc$ and
$T_{1216}=0.80\pm 0.07$ after removing the intrinsic absorption line
at rest frame Ly$\alpha$.

\begin{figure*}
\centering
\begin{minipage}{180mm}
\begin{center}

\psfig{figure=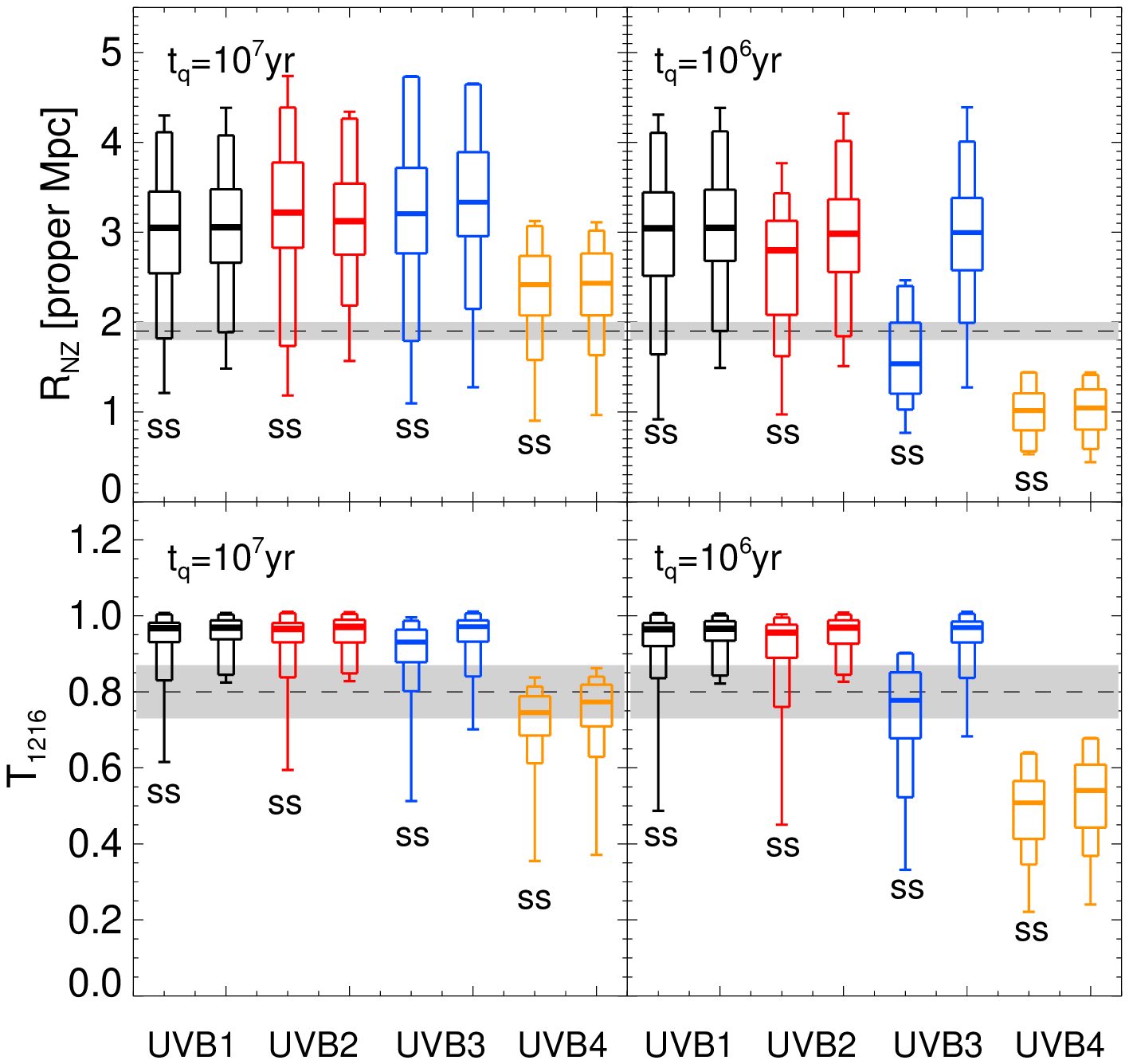,width=0.425\textwidth}
\psfig{figure=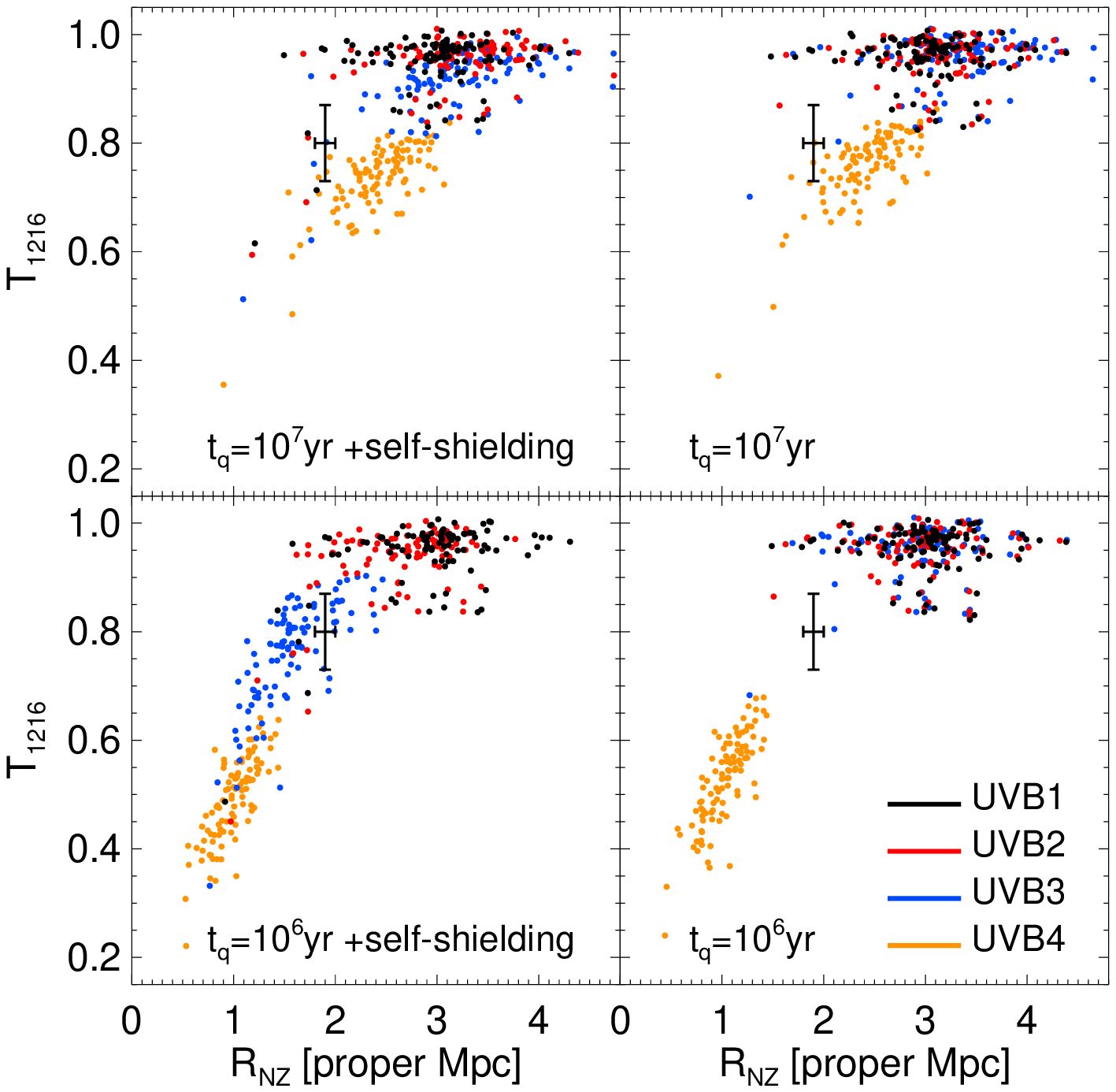,width=0.425\textwidth}

\vspace{-0.4cm}
\caption{{\it Left:} Box-whisker plots of the \Lya near zone size
  (upper panels) and transmitted fraction at the \Lya rest frame
  wavelength (lower panels) measured from synthetic quasar spectra.
  For each UVB model the distributions are shown with (SS) and without
  the self-shielding prescription. The thick horizontal line in each
  box displays the median of the distribution, and the vertical extent
  of the broad and narrow boxes correspond to 68 and 95 per cent of
  the distribution around the median, respectively.  The whiskers
  display the range of the simulated data. The horizontal dashed line
  and grey shaded region in each panel displays the measurement for
  ULAS J1120$+$0641. {\it Right:} Scatter plots of the transmission at
  the \Lya rest frame wavelength, $T_{1216}$, against near zone size,
  $R_{\rm NZ}$.  The data point with error bars corresponds to ULAS
  J1120$+$0641.}

\label{fig:box_whisker}
\end{center}
\end{minipage}
\end{figure*}

Fig.~\ref{fig:box_whisker} displays box-whisker and scatter plots of
the distribution of these two quantities.  Both quantities not only
depend on the assumed UVB model but also on how long the quasar has
been emitting ionising photons.  For example, if self-shielding is
included and the quasar has been bright for only $10^{6}$ yr, both
quantities for UVB3 are strongly affected by neutral gas in Lyman
limit systems (LLSs).  Note also the extended tail in the distribution
of $R_{\rm NZ}$ and $T_{1216}$ when self-shielding is included for
UVB1 and UVB2.  This arises due to the stochastic occurrence of LLSs
and DLAs close to the quasar, and is more prominent the shorter the
time the quasar has been bright and thus able to photo-ionise
proximate optically thick absorbers.  Fig.~\ref{fig:box_whisker}
suggests that the near zone around ULAS J1120$+$0641 is due to either
(i) a partially (fully) neutral surrounding IGM with $\langle f_{\rm
  HI} \rangle_{\rm V}\sim 0.1$ (1.0) and a quasar bright phase of
$\sim 10^{6}\rm\,yr$ ($10^{7}\rm\,yr$) or (ii) the presence of a high
column density \HI system in close proximity to the quasar.  It is
important to note, however, that (i) and (ii) {\it are not mutually
  exclusive}. In our modelling an abundance of optically thick
absorbers close to a quasar is a generic prediction due to the shorter
ionising photon mean free path expected when the IGM is only partially
ionised.  Our simulations confirm the suggestion of M11 that the
spectrum of ULAS J1120$+$0641 is consistent with an already very
highly ionised IGM ($\langle f_{\rm HI} \rangle_{\rm V} \sim
10^{-3}$--$10^{-4}$) if a DLA lies close to the quasar.  However, this
occurs in only $\sim 5$ per cent of our sight-lines for $t_{\rm
  q}=10^{6}$--$10^{7}$ yr.  A fully neutral IGM for $t_{\rm
  q}=10^{6}\rm\,yr$ is ruled out by the near zone size, but is
consistent with the observed transmission for a bright phase duration
of $10^{7}\rm\,yr$.  However, ionisation by lower luminosity sources
expected to cluster around a massive quasar host halo
(\citealt{Lidz07,Wyithe08}) may make this extreme possibility less
likely in practice.

We should caution, however, that if we have underestimated the number
of proximate DLAs present at $z\simeq 7$, if the ionising spectrum of
the quasar is much softer than assumed here, or if the environment of
the quasar host halo is significantly overdense out to $2\rm\, Mpc$
from the quasar (but see \citealt{Calverley11}), then we will
overestimate the likelihood of a significantly neutral surrounding IGM
and/or underestimate the age of the quasar.

\vspace{-0.5cm}
\section{Discussion and Conclusions}  \label{sec:discuss}

We have presented a large suite of realistic synthetic transmission
spectra which model a variety of evolutionary histories for the
ionisation state of the IGM surrounding the quasar ULAS
J1120$+$0641. The transmission profiles depend on the age of the
quasar, the photo-ionisation rate in the surrounding IGM and the
presence of high column density absorbers arising from collapsed, high
density regions.  For the same initial conditions for the IGM
ionisation state the \Lya transmission profile varies greatly due to
the stochasticity of the optically thick absorbers and IGM density
distribution along different sight-lines.

Inferences regarding the ionisation state of the surrounding IGM for
\Lya transmission around ULAS J1120$+$0641 are correlated with the
duration of the optically bright phase of the quasar.  For a short
bright phase duration of $10^{6}\rm \,yr$, the data are consistent
either with an IGM with a volume averaged neutral fraction $\langle
f_{\rm HI}\rangle_{\rm V}\sim 0.1$ or a highly ionised IGM with
$\langle f_{\rm HI} \rangle_{\rm V}\sim 10^{-3}$--$10^{-4}$ and a
proximate DLA.  On the other hand, a bright phase of $10^{7}\rm\,yr$
is consistent with both a highly ionised IGM and proximate DLA or a
fully neutral IGM.  However, we find only $\sim 5$ per cent of our
simulated sight-lines exhibit a proximate DLA capable of producing a
red damping wing, implying the former scenario is less likely (but see
also the caveats in section 4).  In either case, this suggests the
quasar has not been bright for significantly more than $\sim
10^{7}\rm\,yr$; a longer bright phase will produce a near-zone which
is too large, either by enlarging the quasar \HII region or
photo-ionising progressively more proximate optically thick absorbers.
A timescale of $10^{7}\rm\,yr$ is still a factor of five less than the
e-folding time of a black hole growing at the Eddington rate and
radiating with 10 per cent efficiency, as might be expected for a
moderately spinning black hole. For a rapidly spinning black hole the
discrepancy would be even larger (\citealt{Sijacki09}).  Unless the
black hole grows in an earlier optically obscured phase, this adds to
the challenge of building up a black hole with an estimated mass of
$2\times 10^{9}\rm\,M_{\odot}$ so early in the history of the
Universe.  A higher quality spectrum will hopefully exclude or confirm
the possible presence of a proximate DLA by enabling the
identification of associated metal lines and aid in narrowing
constraints on the neutral fraction of the surrounding IGM.

More general conclusions regarding the ionisation state of the IGM at
$z\sim7$ are speculative due to the large fluctuations in the
ionisation state expected on scales much larger than probed by the
near zone of ULAS J1120$+$0641 (\citealt{MesingerFurlanetto08}). If
the neutral fraction in the general IGM at $z\sim 7 $ is indeed
$\langle f_{\rm HI}\rangle_{\rm V}\sim 0.1$ this is nevertheless
consistent with a constant or weakly decreasing ionising emissivity
between $z\sim 6-7$.  The rapid change in the photo-ionisation rate by
two orders of magnitude or more in our models is balanced by a similar
change in the number of LLSs and thus the mean free path of ionising
photons ({\it e.g.} \citealt{McQuinn11}).  Since a constant or perhaps
even rising emissivity is required for the completion of reionisation
at $z\ga 6$ (\citealt{MiraldaEscude03,BoltonHaehnelt07b}), this may be
compatible with the possibility that reionisation was yet to fully
complete by $z=6$ (\citealt{Mesinger10}).  Lastly, the possibility of
a significant neutral fraction in the IGM surrounding ULAS
J1120$+$0641 make it an excellent target for searching for extended
\Lya emission from the expanding ionisation front
(\citealt{Cantalupo08}) as well as studies of the IGM ionisation state
using the redshifted $21\rm\,cm$ transition.


\vspace{-0.74cm}
\begin{thebibliography}{99}
\vspace{-0.1cm}

\bibitem[{{Alvarez} \& {Abel}(2007)}]{AlvarezAbel07}
{Alvarez}, M.~A. \& {Abel}, T. 2007, \mnras, 380, L30

\bibitem[{{Becker} {et~al.}(2007){Becker}, {Rauch}, \& {Sargent}}]{Becker07}
{Becker}, G.~D., {Rauch}, M., \& {Sargent}, W.~L.~W. 2007, \apj, 662, 72

\bibitem[{{Bolton} {et~al.}(2010){Bolton}, {Becker}, {Wyithe}, {Haehnelt}, \&
  {Sargent}}]{Bolton10}
{Bolton}, J.~S., {Becker}, G.~D., {Wyithe}, J.~S.~B., {Haehnelt}, M.~G., \&
  {Sargent}, W.~L.~W. 2010, \mnras, 771

\bibitem[{{Bolton} \& {Haehnelt}(2007{\natexlab{a}})}]{BoltonHaehnelt07}
{Bolton}, J.~S. \& {Haehnelt}, M.~G. 2007{\natexlab{a}}, \mnras, 374, 493

\bibitem[{{Bolton} \& {Haehnelt}(2007{\natexlab{b}})}]{BoltonHaehnelt07b}
{Bolton}, J.~S. \& {Haehnelt}, M.~G. 2007{\natexlab{b}}, \mnras, 382, 325

\bibitem[{{Calverley} {et~al.}(2011){Calverley}, {Becker}, {Haehnelt}, \&
  {Bolton}}]{Calverley11}
{Calverley}, A.~P., {Becker}, G.~D., {Haehnelt}, M.~G., \& {Bolton}, J.~S.
  2011, 412, 2543

\bibitem[{{Cantalupo} {et~al.}(2008){Cantalupo}, {Porciani}, \&
  {Lilly}}]{Cantalupo08}
{Cantalupo}, S., {Porciani}, C., \& {Lilly}, S.~J. 2008, \apj, 672, 48

\bibitem[{{Carilli} {et~al.}(2010){Carilli}, {Wang}, {Fan}, {Walter}, {Kurk},
  {Riechers}, {Wagg}, {Hennawi}, {Jiang}, {Menten}, {Bertoldi}, {Strauss}, \&
  {Cox}}]{Carilli10}
{Carilli}, C.~L. et al. 2010, \apj, 714, 834

\bibitem[{{Fan} {et~al.}(2006){Fan}, {Strauss}, {Richards}, {Hennawi},
  {Becker}, {White}, {Diamond-Stanic}, {Donley}, {Jiang}, {Kim}, {Vestergaard},
  {Young}, {Gunn}, {Lupton}, {Knapp}, {Schneider}, {Brandt}, {Bahcall},
  {Barentine}, {Brinkmann}, {Brewington}, {Fukugita}, {Harvanek}, {Kleinman},
  {Krzesinski}, {Long}, {Neilsen}, {Nitta}, {Snedden}, \& {Voges}}]{Fan06}
{Fan}, X. et al. 2006, \aj, 131, 1203

\bibitem[{{Gunn} \& {Peterson}(1965)}]{GunnPeterson65}
{Gunn}, J.~E. \& {Peterson}, B.~A. 1965, \apj, 142, 1633

\bibitem[{{Lawrence} {et~al.}(2007)}]{Lawrence07}
{Lawrence}, A. et al. 2007, \mnras, 379, 1599

\bibitem[{{Lidz} {et~al.}(2007){Lidz}, {McQuinn}, {Zaldarriaga}, {Hernquist},
  \& {Dutta}}]{Lidz07}
{Lidz}, A., {McQuinn}, M., {Zaldarriaga}, M., {Hernquist}, L., \& {Dutta}, S.
  2007, \apj, 670, 39

\bibitem[{{Maselli} {et~al.}(2009){Maselli}, {Ferrara}, \&
  {Gallerani}}]{Maselli09}
{Maselli}, A., {Ferrara}, A., \& {Gallerani}, S. 2009, \mnras, 395, 1925

\bibitem[{{Maselli} {et~al.}(2007){Maselli}, {Gallerani}, {Ferrara}, \&
  {Choudhury}}]{Maselli07}
{Maselli}, A., {Gallerani}, S., {Ferrara}, A., \& {Choudhury}, T.~R. 2007,
  \mnras, 376, L34

\bibitem[{{McQuinn} {et~al.}(2011){McQuinn}, {Oh}, \&
  {Faucher-Giguere}}]{McQuinn11}
{McQuinn}, M., {Oh}, S.~P., \& {Faucher-Giguere}, C. 2011, \apj submitted (arXiv:1101.1964)

\bibitem[{{Mesinger}(2010)}]{Mesinger10}
{Mesinger}, A. 2010, \mnras, 407, 1328

\bibitem[{{Mesinger} \& {Furlanetto}(2008)}]{MesingerFurlanetto08}
{Mesinger}, A. \& {Furlanetto}, S.~R. 2008, \mnras, 385, 1348

\bibitem[{{Mesinger} \& {Haiman}(2004)}]{MesingerHaiman04}
{Mesinger}, A. \& {Haiman}, Z. 2004, \apjl, 611, L69

\bibitem[{{Miralda-Escude} \& {Rees}(1998)}]{MiraldaRees98}
{Miralda-Escude}, J. \& {Rees}, M.~J. 1998, \apj, 497, 21

\bibitem[{{Miralda-Escud{\'e}}(2003)}]{MiraldaEscude03}
{Miralda-Escud{\'e}}, J. 2003, \apj, 597, 66

\bibitem[{{Mortlock} {et~al.}(2011)}]{Mortlock11}
{Mortlock}, D.J. et al. 2011, \nat, in press

\bibitem[{{Schaye}(2001)}]{Schaye01}
{Schaye}, J. 2001, \apj, 559, 507

\bibitem[{{Sijacki} {et~al.}(2009){Sijacki}, {Springel}, \& {Haehnelt}}]{Sijacki09} 
{Sijacki}. D., {Springel}, V., {Haehnelt} M.~G. 2009,  MNRAS, 400, 100 

\bibitem[{{Springel}(2005)}]{Springel05}
{Springel}, V. 2005, \mnras, 364, 1105

\bibitem[{{Willott} {et~al.}(2010){Willott}, {Albert}, {Arzoumanian},
  {Bergeron}, {Crampton}, {Delorme}, {Hutchings}, {Omont}, {Reyl{\'e}}, \&
  {Schade}}]{Willott10}
{Willott}, C.~J. et al. 2010, \aj, 140, 546

\bibitem[{{Wyithe} {et~al.}(2008){Wyithe}, {Bolton}, \& {Haehnelt}}]{Wyithe08}
{Wyithe}, J.~S.~B., {Bolton}, J.~S., \& {Haehnelt}, M.~G. 2008, \mnras, 383,
  691

\bibitem[{{Wyithe} \& {Loeb}(2004)}]{WyitheLoeb04}
{Wyithe}, J.~S.~B. \& {Loeb}, A. 2004, \nat, 432, 194

\bibitem[{{Wyithe} \& {Bolton}(2011)}]{WyitheBolton11}
{Wyithe}, J.S.B. \& {Bolton}, J.~S. 2011, \mnras, 412, 1926


\end{thebibliography}
\end{document}